\documentclass[journal]{IEEEtran}

\usepackage{amsmath,graphicx}
\usepackage{amsmath, amssymb, amsfonts, amsbsy, mathrsfs, bm}
\usepackage{cite, enumerate}
\usepackage[table]{xcolor}
\usepackage[normalem]{ulem}
\usepackage{dsfont}

\newtheorem{assumption}{Assumption}

\def\diag{{\mathrm{diag}}}

\newcommand{\E}[1]{\mathbb{E}\left[ #1 \right]}

\def\bw{{\pmb{w}}}
\def\bq{{\pmb{q}}}
\def\bu{{\pmb{u}}}
\def\bd{{\pmb{d}}}
\def\bv{{\pmb{v}}}
\def\bh{{\pmb{h}}}

\title{Steady-State Performance of Incremental LMS Strategies for Parameter Estimation Over Fading Wireless Channels}
\author{Azam~Khalili,~and~ %
        Amir~Rastegarnia
       \thanks{A. Khalili and A. Rastegarnia  are with  Department of Electrical Engineering, Malayer University, Malayer 65719-95863, Iran e-mails: (rastegar,tinati,a-khalili@tabrizu.ac.ir).}}

\begin{document}
%
\maketitle
\begin{abstract}
We study the effect of fading in the communication channels between nodes on the performance of the incremental least mean square (ILMS) algorithm. We derive steady-state performance metrics, including the mean-square deviation (MSD), excess mean-square error (EMSE), and mean-square error (MSE). We obtain the sufficient conditions to ensure mean-square convergence, and verify our results through simulations. Simulation results show that our theoretical analysis closely matches the actual steady state performance.
\end{abstract}

\section{Introduction}
\label{sec:intro}
There are already several useful strategies for estimation over distributed networks. An example of a distributed method is the consensus strategy \cite{Boyd06, Schizas09,Kar11} in which each node performs a local estimation and fuses its estimate with those of its neighbors so that all nodes converge to the same estimate as the number of iterations increases. The main problem of mentioned methods is that the framework does not allow the network to undertake a continuous learning and optimization \cite{chen12} which motivated the development of adaptive networks. An adaptive  network is a collection of agents (nodes) that collaborate with each other through in-network local processing rules in order to estimate and track parameters of interest  \cite{sayed13b}. Two major classes of adaptive networks are incremental strategy \cite{lopes07, sayed06, tak08, catt11b} and diffusion strategy \cite{lopes08, catt08,catt09}. In the incremental mode, nodes passing updates to each other in a Hamiltonian cycle in the network, while in diffusion mode, each node can communicate with a subset of neighboring nodes.

In this paper, we investigate the performance of the ILMS algorithm in a wireless adaptive network with communication links between neighboring nodes modeled as fading channels. In the original ILMS strategy proposed by \cite{lopes06}, it is assumed that nodes communicate with each other via ideal links which is typically not true in practice. In \cite{azam11a,azam11b,azam12a,azam12b,zhao12a}, the effect of additive link noise in the communication channels between nodes have been investigated. In \cite{zhao12}, the performance of general adaptive diffusion algorithms in the presence of imperfect information exchanges, including quantization errors, and model non-stationarities has been considered. All these works however can not be directly applied to a wireless adaptive network, whose communication links are modeled by fading channels \cite{Tse05}. The references \cite{Abdolee11,Abdolee13} propose diffusion LMS algorithms for wireless sensor networks with fading channels but under the  assumption that channel state information is known so that channel equalization can be performed. 

In this paper, our objective is to investigate how the performance of the ILMS algorithm is affected by the channel fading statistics of the communication channels between nodes. We show that the ILMS algorithm over fading channels, like the traditional ILMS algorithm, is stable in mean if the step size is chosen in an appropriate range. This range now not only depends on the regression correlation, but also on the mean of the channel fading gains. Moreover, our analysis reveals that in general, fading communication channels lead to biased estimates in steady state. 
	 We derive closed-form expressions for the steady state performance metrics, including the mean-square deviation (MSD), excess mean-square error (EMSE) and mean-square error (MSE), of the ILMS algorithm in the presence of fading channels, and under a Gaussian model. We show explicitly how these metrics are affected by the fading channel statistics. 	 We derive sufficient conditions for the convergence of the MSD, EMSE and MSE, and show that for a fixed step size, mean-square stability is lost if the channel gain variances become large. We present simulations to verify that our theoretical analysis closely matches the actual steady state performance.

\textit{Notation}: We adopt boldface letters for random quantities. The symbol $^*$ denotes conjugation for scalars and Hermitian transpose for matrices. The notation $\rm{diag}\{\cdot\}$ will be used in two ways: $X=\mathrm{diag}\{x\}$ is a diagonal matrix whose entries are those of the vector $x$, and $x=\rm{diag}\{X\}$ is a vector containing the main diagonal of $X$. If $\Sigma$ is a matrix, we use the notation $\| x\|_{\Sigma}^2=x^* \Sigma x$ for the weighted square norm of $x$. If $\sigma$ is a vector, the notation $\| x\|_{\sigma}$ is used to represent $\| x \|_{\mathrm{diag}\{\sigma\}}$.

\section{ILMS Algorithm over fading channels}\label{sect:ILMS}
Consider a network composed of $N$ nodes. At time $i$, node $k$ observes a scalar measurement $\bd_k(i)$ and a $1\times M$ regression vector $\bu_{k,i}$,  which are related via a linear regression model
\begin{equation} \label{dmodel}
\bd_k(i)=\bu_{k,i} w^o+\bv_k(i),
\end{equation}
where $\bv_k(i)$ is the observation (or measurement) noise, and the $M \times 1$ vector $w^o$ is a unknown  vector.  The goal of the network is to estimate $w^o$, at every node $k$, using all observed  data in the entire network. In the incremental LMS algorithm \cite{lopes06, lopes07}, each node $k$ receives a local estimate from the previous node $k-1$, updates it using its local data, and then sends it to the next node $k+1$. The update equations for the ILMS algorithm, at iteration $i$ is given by \cite{lopes07} 
\begin{equation} \label{ilms}
\left\{ \begin{array}{l}
 \bw _{0,i}  \leftarrow \pmb{w}_{N, i - 1}  \\ 
 \bw _{k,i}  = \mathop {\bw _{k - 1,i}  + \mu_k } \bu_{k,i}^* ( {\bd_k (i) - \bu_{k,i} \bw _{k - 1,i} })\, \\ 
 \end{array} \right.
\end{equation}
where $\bw_{k,i}$ is the local estimate of the node $k$ at time $i$. It is shown in \cite{lopes07} that  ${\bw_{k,i}} \to {w^o}$ as $i \rightarrow \infty$ for every node $k$.

Now, we consider the case where nodes communicate over fading channels. By incorporating the impact of fading channels, the update equation for node $k$ in \eqref{ilms} becomes
\begin{align} 
\bw_{k,i} & = \pmb{h}_k(i) \bw_{k - 1,i} + {{\pmb{q}}_{k,i}} \nonumber \\ 
&\quad  + {\mu _k}{\pmb{u}}_{k,i}^ * \left({{\pmb{d}}_k}(i) - {{\pmb{u}}_{k,i}}({\pmb{h}_k}(i) \bw _{k - 1,i}+ {{\bq}_{k,i}}) \right), \label{films}
\end{align}
where ${\pmb{h}_k}(i)$ is the channel gain at time $i$ for the communication channel between node $k-1$ and $k$, and $\bq_{k,i}$ is the zero mean channel noise with covariance matrix $Q_k=\mathbb{E}[\bq_{k,i}\bq_{k,i}^*]$. We assume phase coherent reception at every node $k$ and model the channel gain as a non-negative random variable. We make the following assumptions regarding the fading channel statistics.
\begin{assumption}\label{assumpt:fading}\
\begin{enumerate}
	\item The channel gains $\pmb{h}_k(i)$ for all nodes $k=1,\ldots,N$, and all observation times $i \geq 1$, are independent of each other. For each node $k$, the channel gains $\{\bh_k(i) : i \geq 1\}$ are identically distributed. 
	\item The channel gains $\pmb{h}_k(i)$ for all nodes $k=1,\ldots,N$, and all observation times $i \geq 1$, are independent of $(\pmb{d}_l(j), \pmb{u}_{l,j})$ for all $l$ and $j$. 
	\end{enumerate}
\end{assumption}

Let $m_k = \E{\bh_k(i)}$ and $s_k = \E{\bh_k^2(i)}$ be the mean and second order moment of the channel gain for node $k$, respectively. We note that in practice, even if the nodes perform channel state estimation, it is not possible to measure the channel gains with absolute certainty, especially if the channels experience fast fading. In this case, without loss of generality, the channel gain for node $k$ can still be modeled as a non-negative random variable with a non-trivial variance. 

\section{Performance Analysis}\label{sect:Performance}
In this section, we analyze the mean stability and steady state mean-square performance of the ILMS algorithm when communication channels between nodes are fading channels. Our analysis is based on the energy conservation approach of \cite{lopes07}.
We make the following assumptions regarding the data model in \eqref{dmodel}. These assumptions are commonly assumed in the literature \cite{lopes06, sayed13b}.
\begin{assumption}\label{assumpt:model}\
\begin{enumerate}[(i)]
	\item The regression vectors $\bu_{k,i}$ are independent over node indices $k$ and observation times $i$. 
	\item The measurement noises $\bv_k(i)$ are independent of each other and the regression vectors $\bu_{k,i}$.
\end{enumerate}
\end{assumption}

In our analysis, we will use the deviation between an observed measurement and its prediction based on the current local estimate, which is defined as
\begin{align*}
\pmb{e}_k (i) &= \pmb{d}_k (i) - \pmb{u}_{k,i} \bw_{k - 1,i},
\end{align*}
and the weight error vector, which is the deviation between the local estimate $\bw_{k,i}$ and its true value $w^o$, given by
\begin{align*}
\widetilde{\bw}_{k,i} &= w^o- \bw_{k,i}.
\end{align*}
By subtracting $w^o$ from both sides of \eqref{films} and using the definition of $\widetilde{\bw}_{k,i}$ we obtain
\begin{align} \label{wer}
\widetilde{\bw}_{k,i} &= {\pmb{h}_k}(i)\widetilde{\bw} _{k - 1,i}  \nonumber\\
 &+ (1 - {\pmb{h}_k}(i)){w^o} - {\mu _k}{\pmb{h}_k}(i)\pmb{u}_{k,i}^ * {\pmb{u}_{k,i}}\widetilde{\bw} _{k - 1,i} \nonumber\\
 &- {\mu _k}\pmb{u}_{k,i}^ * {\pmb{v}_k}(i) - {\mu _k}(1 - {\pmb{h}_k}(i))\pmb{u}_{k,i}^ * {\pmb{u}_{k,i}}{w^o} - {\pmb{q}_{k,i}} \nonumber\\
 &+ {\mu _k}\pmb{u}_{k,i}^ * {\pmb{u}_{k,i}}{\pmb{q}_{k,i}}.
\end{align}
In the steady-state analysis, we are interested to quantify the performance using the following metrics at every node $k$:
\begin{align} 
\eta_k & \triangleq   \mathop {\lim }\limits_{i \to \infty} \mathbb{E}\left[\|\widetilde{\bw}_{k-1,i}\|^2 \right]\     (\mathrm{MSD})         \label{msd} \\
\zeta _k  & \triangleq    \mathop {\lim }\limits_{i \to \infty} \mathbb{E}\left[ \| {\widetilde{\bw} _{k - 1,i} } \|_{R_{u,k} }^2 \right] \,({\rm{EMSE}} ) \label{emse}  \\
\xi_k & \triangleq   \mathop {\lim }\limits_{i \to \infty} \mathbb{E}\left[|\pmb{e}_{k}(i)|^2\right]= \zeta_k+\sigma_{v,k}^2\ (\mathrm{MSE})     \label{mse}
\end{align}
To derive the above steady state performance metrics, we need to evaluate quantities of the form $\mathbb{E}\left[\| {\widetilde{\bw} _{k,i}}\|_{{\Sigma _k}}^2 \right]$ where $\Sigma_k$ is a positive semi-definite Hermitian matrix. To this end, we consider the weight vector update equation given by \eqref{wer}. Let
\begin{align}
C_{k,i} &=m_k J_k C_{k-1,i}+(1-m_k)J_k ,
\end{align}
where $J_k \triangleq I-\mu_k R_{u,k}$, $C_{0,i} =C_{N,i-1}$, and  $C_{0,1}=I$. Note that the matrix $C_{k,i}$ is such that
\begin{equation}\label{eqn:Ck}
\mathbb{E}\left[\widetilde{\bw}_{k ,i}\right] = {C_{k ,i}}{w^o}.
\end{equation}
By equating the weighted norm of both sides of \eqref{wer}, taking expectations and using Assumptions \ref{assumpt:model} and \ref{assumpt:fading}, and \eqref{eqn:Ck}, we obtain the following recursive relationship:
\begin{align} \label{var}
 \mathbb{E}\left[\| {\widetilde{\bw} _{k,i}}\|_{\Sigma_k}^2 \right]
&= \mathbb{E}\left[ \| {\widetilde{\bw} _{k - 1,i}}\|_{{\Sigma'_k}}^2\right] + \mu _k^2\sigma _{v,k}^2 \mathbb{E}\left[\| {{\pmb{u}_{k,i}}}\|_{{\Sigma _k}}^2\right]  \nonumber\\
 & + \mathbb{E}\left[ \| {{\pmb{q}_{k,i}}}\|_{{G_k}}^2\right]+ \| {{w^o}}\|_{{T_k} + {H_{k,i}}}^2,
\end{align}
where 
\begin{align}
{G_k} &= {\Sigma _k} - {\mu _k} \mathbb{E}\left[{\Sigma _k}\pmb{u}_{k,i}^ * {\pmb{u}_{k,i}} + \pmb{u}_{k,i}^ * {\pmb{u}_{k,i}}{\Sigma _k}\right] \\ \nonumber
&   \hspace{3cm}  + \mu _k^2 \mathbb{E}\left[\| {{\pmb{u}_{k,i}}} \|_{{\Sigma _k}}^2\pmb{u}_{k,i}^ * {\pmb{u}_{k,i}}\right] \\
\Sigma'_k &= s_k {G_k}     \\
{T_k} &= (1-2m_k+s_k) {G_k}      \\
{H_{k,i}} &= (m_k-s_k) ({C_{k - 1,i}}{G_k} + {G_k}{C_{k - 1,i}}).  \label{hre} 
\end{align}
In order to compute all the moments that appear in the recursive equation \eqref{var} and to obtain closed-form expressions, we now make the following assumption regarding the regression vectors $\bu_k$, for all nodes $k=1,\ldots,N$.
\begin{assumption}\label{assumpt:Gaussian}
For each $k=1,\ldots,N$, the distribution of $\bu_k$ is a Gaussian distribution with
\begin{equation} \label{deco}
R_{u,k}= U_k \Lambda _k U_k^ *, 
\end{equation}
where $\Lambda _k $ is a diagonal matrix with diagonal elements being the eigenvalues of the correlation matrix $R_{u,k}$, and $U_k$ is unitary matrix.
\end{assumption}

Making use of Assumption \ref{assumpt:Gaussian}, we further define the following transformed quantities:
\begin{align} 
{{\bar {\bw} }_{k,i}} &= U_k^ * {\widetilde{\bw} _{k,i}},
 {{\bar \Sigma }_k} = U_k^ * {\Sigma _k}{U_k}, \\
{{\bar \Sigma '}_k} &= U_k^ * {{\Sigma '}_k}{U_k}, 
 {{\bar {\pmb{u}}}_{k,i}} = {\pmb{u}_{k,i}}{U_k}, \\
{{\bar T}_k} &= U_k^ * {T_k}{U_k},
 {{\bar H}_{k,i}} = U_k^ * {H_{k,i}}{U_k}, \\
{{\bar w}^o} &= U_k^ * {w^o},
 {{\bar C}_{k - 1, {i}}} = U_k^ * {C_{k - 1, {i}}}{U_k}, \\
{{\bar Q}_k} &= U_k^ * {Q_k}{U_k},
 {D={{\bar w}^o}{{\bar w}^{o * }} = w^o w^{o*}}.
\end{align}

From the above definitions, equation \eqref{var} can now be rewritten in the following equivalent form
\begin{align} \label{tvar}
\mathbb{E} \left[\| {\bar {\bw} _{k,i}} \|_{{{\bar \Sigma }_k}}^2 \right] &= \mathbb{E} \left[\| {\bar {\bw} _{k - 1,i}} \|_{{{\bar \Sigma '}_k}}^2 \right]+ {\mu_k ^2}\sigma _{v,k}^2 \mathbb{E} \left[\| {{{\bar {\pmb{u}}}_{k,i}}} \|_{{{\bar \Sigma }_k}}^2 \right]  \nonumber\\
&   \hspace{1cm} + \mathbb{E}\left[\| {{{\bar {\pmb{q}}}_{k,i}}} \|_{{{\bar G}_k}}^2 \right] + \| {{{\bar w}^o}} \|_{{{\bar T}_k} + {{\bar H}_{k,i}}}^2
\end{align}
where in \eqref{tvar} we have 
\begin{align}
{{\bar G}_k} &= {{\bar \Sigma }_k} - {\mu _k} \mathbb{E}\left[{{\bar \Sigma }_k}\bar {\pmb{u}}_{k,i}^ * {{\bar {\pmb{u}}}_{k,i}} + \bar {\pmb{u}}_{k,i}^ * {{\bar {\pmb{u}}}_{k,i}} {{\bar \Sigma }_k}\right] \nonumber \\
& \hspace{2cm} + \mu _k^2 \mathbb{E} \| {{{\bar {\pmb{u}}}_{k,i}}} \|_{{{\bar \Sigma }_k}}^2\bar {\pmb{u}}_{k,i}^ * {{\bar {\pmb{u}}}_{k,i}} \nonumber 
\\
 &= {{\bar \Sigma }_k} - {\mu _k}({{\bar \Sigma }_k}{\Lambda _k} + {\Lambda _k}{{\bar \Sigma }_k})  \nonumber\\ 
& \hspace{2cm}  + \mu _k^2({\Lambda _k}{\rm{Tr[}}{{\bar \Sigma }_k}{\Lambda _k}]+ \gamma {\Lambda _k}{{\bar \Sigma }_k}{\Lambda _k})    \label{gbar}
\\
{{\bar \Sigma '}_k} &=  s_k {{\bar G}_k}   \label{sbar}
\\
{{\bar T}_k} &=  (1-2m_k+s_k) {{\bar G}_k}    \nonumber
\\
 {\bar H}_{k,i}&=    (m_k-s_k)  ({{\bar C}_{k - 1,i}}{{\bar G}_k} + {{\bar G}_k}{{\bar C}_{k - 1,i}})\nonumber\\ 
&=      2 (m_k-s_k) {{\bar C}_{k - 1,i}}{{\bar G}_k}  \label{tbar} 
\end{align}
Further algebraic manipulations of \eqref{tvar} yields
\begin{align} \label{cvar}
\mathbb{E}\left[ \| {{{\bar {\bw}}_{k,i}}} \|_{{{\bar \Sigma }_k}}^2\right] &= \mathbb{E}\left[ \| {{{\bar {\bw} }_{k - 1,i}}} \|_{{{\bar \Sigma '}_k}}^2 \right]+ \mu _k^2\sigma _{v,k}^2{\rm{Tr[}}{\Lambda _k}{{\bar \Sigma }_k}] \nonumber\\
& + {\rm{Tr[}}{{\bar Q}_k}{{\bar G}_k}] + {\rm{Tr[}}D{{\bar T}_k}] + {\rm{Tr[}}D{{\bar H}_{k,i}}].
\end{align}
To derive \eqref{msd}-\eqref{mse}, we only need to consider the case where $\bar{\Sigma}{_k}$ is a diagonal matrix. In this case, matrix $\bar{\Sigma}^\prime{_k}$  is also a diagonal matrix. We let
\begin{equation} \label{dvec}
\bar{\sigma}_k   \triangleq {\rm{\diag}}\{ \bar{\Sigma}_k  \}, \quad \bar{\sigma}_k ^\prime   \triangleq {\rm{diag}}\{ \bar{\Sigma}_k  ^\prime  \},  \quad \lambda _k  \triangleq {\rm{diag}}\{ \Lambda _k \},
\end{equation}
and
\begin{equation} \label{fbar}
{{\bar F}_k} = I - \mu_k X_k +\mu_k^2 Y_k,
\end{equation}
with $X_k=2 {\Lambda _k}$ and $Y_k=\Lambda _k^2 + {\lambda _k}\lambda _k^T$. The $M \times M$ matrix $\bar F_k$ contains the statistics of data local to node $k$. We then have 
\begin{equation} \label{fvar}
\mathbb{E}\left[ \| {\bar {\bw} _{k,i}} \|_{{{\bar \sigma }_k}}^2 \right] = \mathbb{E}\left[ \| {\bar {\bw} _{k - 1,i}} \|_{{{\bar {\sigma} '}_k}}^2 \right]+ {{g}_{k,i}}{{\bar {\sigma}_k}},
\end{equation}
where $g_{k,i}$ and ${\bar {{\sigma}} '}_k$ are given  respectively by
\begin{align}
g_{k,i} &= \mu _k^2\sigma _{v,k}^2\lambda _k^T + \diag{\{ {{\bar Q}_k}\} ^T}{{\bar F}_k} \nonumber \\
&  \hspace{.1cm}+ (1-2m_k+s_k)  \diag{\{ D\} ^T}{{\bar F}_k} \nonumber \\ 
&  \hspace{.1cm} + 2(m_k-s_k) \diag{\{ D\} ^T}{{\bar C}_{k - 1,i}}{{\bar F}_k}, \nonumber \\
{\bar {{\sigma}} '}_k &=  s_k {{\bar F}_k}{{\bar {{\sigma}} }_k}   \label{sfor}.
\end{align}

We can use \eqref{fvar} to derive conditions that guarantee convergence in the mean-square sense for the ILMS algorithm with fading channels. Under assumptions \ref{assumpt:model}, \ref{assumpt:fading} and \ref{assumpt:Gaussian}, the ILMS algorithm over fading channels converges in the mean-square sense if the step sizes $\mu_k$ are chosen to be sufficiently small so that the 
\begin{align}\label{msconvcond}
s_k \rho(\bar{{F}}_k)  &< 1.
\end{align}
Suppose that $s_k = s$ for all nodes $k$, and the step sizes $\mu_k$ are fixed. Then, if $s$ is sufficiently large, the left hand side of \eqref{fvar} diverges and we no longer have mean-square stability. This shows that deteriorating fading conditions have detrimental impact on the ILMS algorithm, and care should be taken to adjust the step sizes according to \eqref{msconvcond}.  

Assuming that step sizes are chosen sufficiently small, and by letting $i\to\infty$, the recursive equation \eqref{fvar} at steady-state gives
\begin{equation} \label{fvar_steady}
\mathbb{E}\left[ \| {\bar {\bw} _{k,\infty}} \|_{{{\bar \sigma }_k}}^2 \right] = \mathbb{E}\left[ \| {\bar {\bw} _{k - 1,\infty}} \|_{{{\bar {\sigma} '}_k}}^2 \right]+ {{g}_k}{{\bar {\sigma}_k}},
\end{equation}
where 
\begin{align}
g_{k} &= \mu _k^2\sigma _{v,k}^2\lambda _k^T + \diag{\{ {{\bar Q}_k}\} ^T}{{\bar F}_k} \nonumber \\
&  \hspace{.1cm}+ (1-2m_k+s_k)  \diag{\{ D\} ^T}{{\bar F}_k} \nonumber \\ 
&  \hspace{.1cm} + 2(m_k-s_k) \diag{\{ D\} ^T}{{\bar C}_{k - 1,\infty}}{{\bar F}_k}, \label{gfor}
\end{align}
Moreover, ${{\bar C}_{k - 1,\infty}}$ in \eqref{gfor} is given by
\begin{align} \label{Ckinf}
{{\bar C}_{k - 1,\infty}} &=U_k^* \Bigg( (I-\mathcal{M})^{-1} \times \nonumber \\
&  \hspace{1.2cm} \sum_{n=1}^{N} \bigg((1-m_n) J_n  \prod_{\ell=n+1}^{N} m_{\ell} J_{\ell} \bigg) \Bigg) U_k.
\end{align}

We observe that \eqref{fvar_steady} shows how $\mathbb{E}\left[ \| {\bar {\bw} _{k,\infty}} \|_{{{\bar \sigma }_k}}^2 \right]$ evolves through the network, which in its current form makes it difficult to derive the desired metrics \eqref{msd}-\eqref{mse} directly. In fact, we have to find a recursive equation that reveals how $\mathbb{E}\left[ \| {\bar {\bw} _{k,i}} \|_{{{\bar \sigma }_k}}^2 \right]$ evolves in time. By iterating \eqref{fvar}, and using $\bw_{0,i+1}= \bw _{N,i}$, we can obtain a set of $N$ coupled equations. With suitable manipulation of these equations, along with proper selections of ${{\bar {\sigma}}_k}$, it is possible to solve the resulting equalities to derive the desired metrics. Following the argument given in \cite{lopes07}, we can derive the required metrics in a similar way as 
\begin{align}
\eta _k  &= a_k (I - \Pi _{k,1} )^{ - 1} \mathds{1}  \label{fmsd} \\ 
 \zeta _k  &= a_k (I - \Pi _{k,1} )^{ - 1} \lambda _k   \label{femse} \\
 \xi _k  &= \zeta _k  + \sigma _{v,k}^2   \label{fmse}
\end{align}
where 
\begin{align} \label{pfor}
\Pi _{k,l}  & \triangleq \bigg(\prod_{k=1}^{N} s_k\bigg) \bigg(\bar{F} _{k + l - 1}  \bar{F} _{k + l}  \cdots  \bar{F} _N   \bar{F} _1  \cdots \bar{F} _{k - 1}\bigg), \\
a_k  & \triangleq g_k \Pi _{k,2}  + g_{k + 1} \Pi _{k,3}  +  \ldots  + g_{k - 2} \Pi _{k,N}  + g_{k - 1}, \label{afor}
\end{align}
where $l = 1, \cdots ,N$ and all the subscripts are in $\bmod \;N$.

\section{Simulation Results}\label{sect:Simulation}
\begin{figure}[t]
\centering 
\includegraphics [width=7cm]{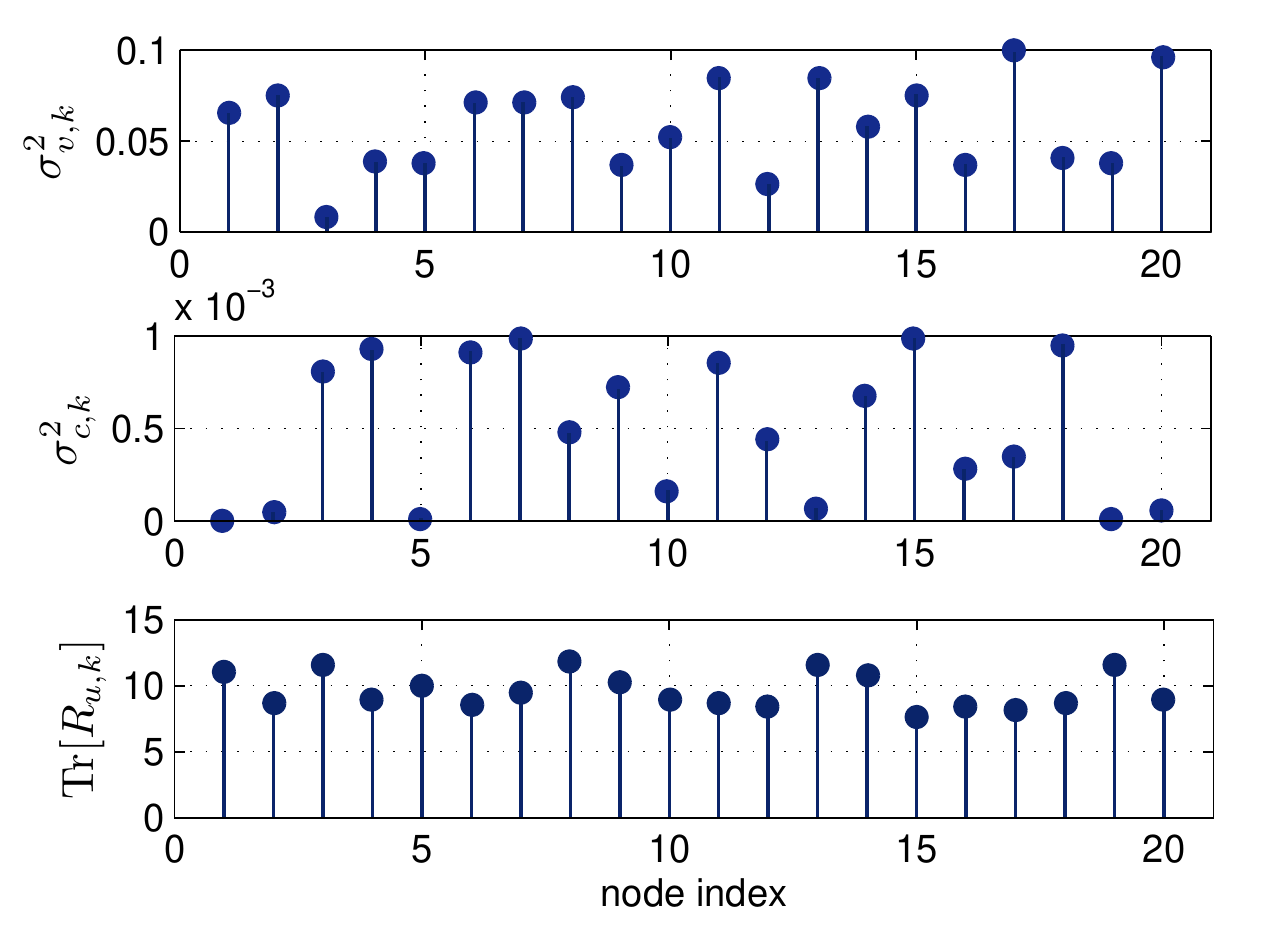} 
\centering \caption{Node profile and channel noise information: $\sigma_{v,k}^2$ (up), $\sigma_{c,k}^2$ (middle) and $\mathrm{Tr}(R_{u,k})$ (down).}
\label{fig:prof}
\end{figure}

We illustrate the results via simulations. We assume a network composed of $N=20$ nodes,  where the nodes are connected via a ring topology as in the ILMS algorithm. The regressors $\bu_{k,i}$ are generated as independent realizations of a Gaussian distribution with a covariance matrix $R_{u,k}$ whose eigenvalue spread is 5. The measurement data $\bd_k(i)$ at each node $k$ is generated by using the data model \eqref{dmodel} where the parameter $w^o$ is chosen to be $[1~1~1~1]^T/2$, and the observation noise $\bv_k(i)$ is drawn from a Gaussian distribution with variance $\sigma_{v,k}^2$ as shown in Figure \ref{fig:prof}.  The additive channel noises are generated from Gaussian distributions with covariance matrix $Q_k=\sigma_{c,k}^2 I$, for $k=1,\ldots,20$. The values of $\sigma_{c,k}^2$ are shown in Figure \ref{fig:prof}. We generate the channel gains $\bh_k(i)$ using a Rayleigh distribution with $m_k=\sqrt{2}/2$ for all values of $k$. To obtain the steady-state values of MSD, EMSE and MSE, we run the ILMS algorithm with 2000 iterations and average the last 200 samples. Finally, each steady-state value is obtained by averaging over 100 independent runs. 

\begin{figure}[th]
\centering 
\includegraphics [width=7cm]{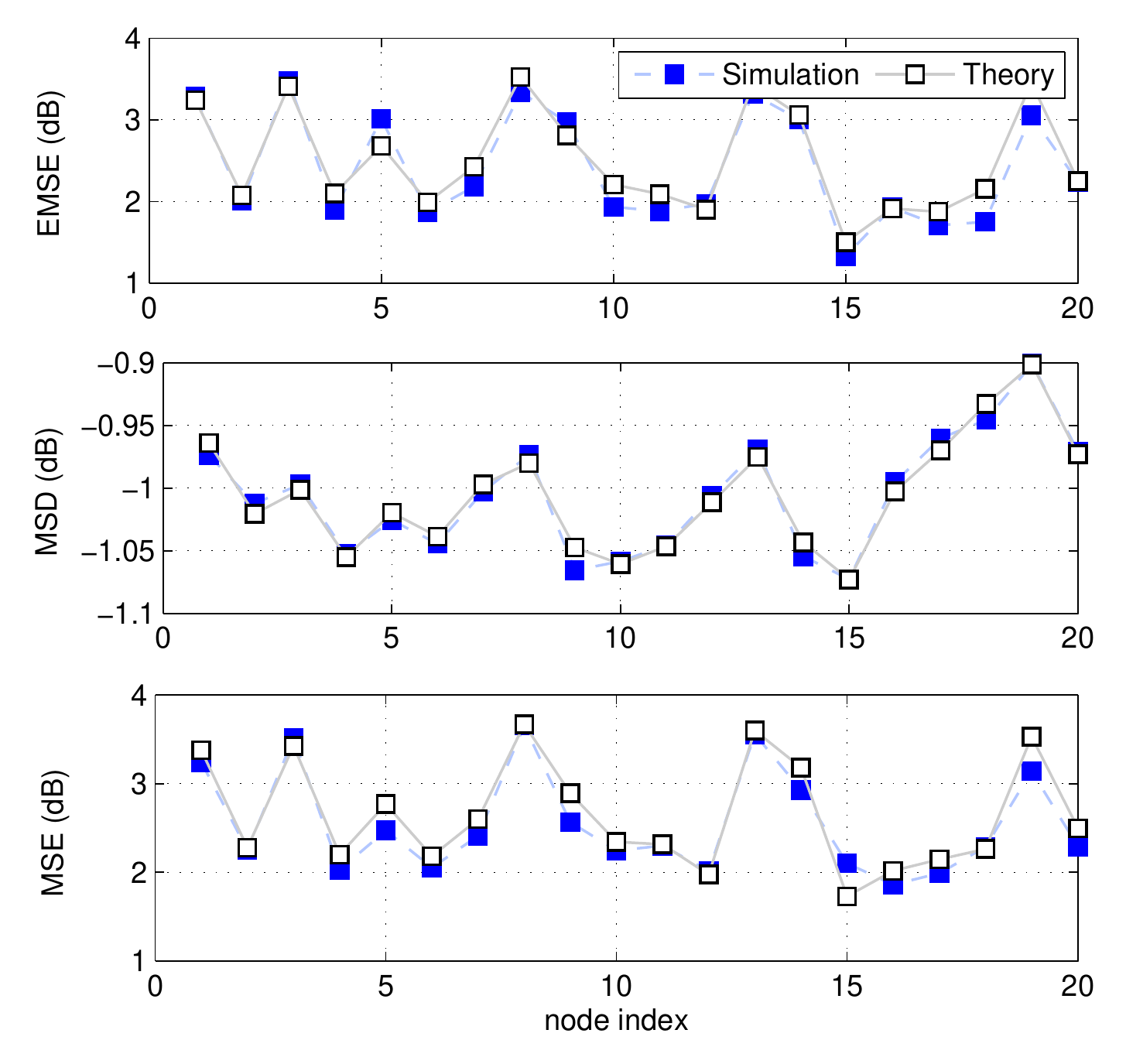} 
\centering \caption{Steady-state curves versus for each individual node $k$, $\mu=0.02$.}
\label{fig:std}
\end{figure}

In Figure \ref{fig:std}, we show the steady-state performance metrics MSD, EMSE and MSE as functions of the node index $k$ when the step size $\mu=0.02$. We can see that the simulated results closely match the theoretical results. 



\section{Conclusion}\label{sect:Conclusion}
In this paper, we have investigated the steady state performance of the ILMS algorithm when the links between nodes are fading channels, and we do not have perfect channel state information. Our analysis reveals how the mean-square stability depends on the channel gain variances. We also derived steady state performance metrics, including the MSD, EMSE and MSE. We present simulation results to verify our theoretical analysis.


\end{document}